# Investigating individual arsenic dopant atoms in silicon using low-temperature scanning tunnelling microscopy

Kitiphat Sinthiptharakoon[1,2], Steven R. Schofield[1,3], Philipp Studer[1,2], Veronika Brázdová[1], Cyrus F. Hirjibehedin[1,3,4], David R. Bowler[1,3] and Neil J. Curson[1,2]

[1] London Centre for Nanotechnology, UCL, United Kingdom, WC1H 0AH

[2] Department of Electronic and Electrical Engineering, UCL, United Kingdom, WC1E 7JE

[3] Department of Physics and Astronomy, UCL, United Kingdom, WC1E 6BT

[4] Department of Chemistry, UCL, United Kingdom, WC1H 0AJ

E-mail: n.curson@ucl.ac.uk

**Abstract**

We study sub-surface arsenic dopants in a hydrogen terminated Si(001) sample at 77 K, using scanning tunnelling microscopy and spectroscopy. We observe a number of different dopant related features that fall into two classes, which we call As1 and As2. When imaged in occupied states the As1 features appear as anisotropic protrusions superimposed on the silicon surface topography, and have maximum intensities lying along particular crystallographic orientations. In empty-state images the features all exhibit long-range circular protrusions. The images are consistent with buried dopants that are in the electrically neutral ($D^0$) charge state when imaged in filled states, but become positively charged ($D^+$) through electrostatic ionisation when imaged under empty state conditions, similar to previous observations of acceptors in GaAs. Density functional theory (DFT) calculations predict that As dopants in the third layer of the sample induce two states lying just below the conduction band edge, which hybridize with the surface structure creating features with the surface symmetry consistent with our STM images. The As2 features have the surprising characteristic of appearing as a protrusion in filled state images and an isotropic depression in empty state images, suggesting they are negatively charged at all biases. We discuss the possible origins of this feature.

## 1. Introduction

The study of dopants in silicon has been of scientific interest and technological importance for many decades. In recent years, there has been strong interest in measuring the properties of individual dopants due to the continued miniaturization of integrated circuit components, and the development of quantum information processing. For example, phosphorous atoms in silicon are seen as potential building blocks for a quantum computer, whose quantum bits may be comprised of the electron [1] or nuclear spins [2] of donor atoms. These proposals were recently made even more promising by the measurement of very long spin lifetimes of P in Si [3]. The capability of fabricating atomic-scale devices by deterministically placing individual dopant atoms in silicon with scanning tunnelling microscopy (STM), is well established [4-6]. When combined with scanning tunnelling spectroscopy (STS) this technique provides the capability for measuring the structural and electronic properties of substitutional dopant atoms in semiconductors via the ability to map the local density of states (LDOS) at the atomic scale.

Substitutional dopants have been studied with STM/STS in a variety of semiconductors, with the majority of work concentrating on the cleaved GaAs(110) surface. In general, the appearance of such dopants in STM images is dependent on whether they are donors or acceptors, and whether their intrinsic electronic states are shallow (~<50 meV) or deep (~> 50 meV). Shallow donors in GaAs, such as Si [7] and Te [8], are always imaged as isotropic circular protrusions surrounded by a dark halo in low-bias filled-state images, and dark depressions in high-bias filled-state STM images. This has been interpreted as imaging the screened Coulomb potential of the ionized donors [7]. In contrast, shallow acceptors in GaAs, such as Zn [9-15], Cd [10], Be [11], and C [12], were seen to induce anisotropic triangular-shaped protrusions at low positive sample bias (empty state), and dark isotropic depressions in high-bias empty-state images. These triangular dopant features were first observed nearly two decades ago [9]; however, the origin of their triangular shape in empty-state images remains controversial. It has been suggested that their characteristic appearance results from a direct image of the acceptor ground state [9, 10]; tunnelling through an excited state of the acceptor that originates from the highly anisotropic heavy hole band [11]; or resonant tunnelling through evanescent gap states [13, 14]. Notwithstanding these varied interpretations, most agree that the occurrence of the triangular imaging contrast coincides with a tunnelling regime where tunnelling into the conduction band is largely suppressed [15]. In case of energetically deep acceptors such as Mn in GaAs, a cross-like protrusion feature was observed in low-bias empty-state images, which is believed to arise due to imaging the wavefunction of neutral Mn acceptor [16]. Generally, the features induced by subsurface dopant atoms become less pronounced at large biases in both filled- and empty-state images due to the dominance of tunnelling currents arising from the valence and conduction bands, respectively.

The STM/STS observation of dopant atoms in silicon has received less attention in the literature, and to date there have been no reports of imaging dopant wavefunctions. The observation of circularly symmetric depressions and protrusions associated with screened Coulomb potentials resulting from ionized dopants have been reported beneath both clean [17, 18] and hydrogen-terminated [19-21] Si(001) surfaces. These studies examined As [20] and P [17, 21] donors and B acceptors [18-20].

Here, we present STM/STS observations of individual As dopant atoms beneath Si(001):H at 77K. We report for the first time the observation of anisotropic protrusions for dopants atoms in silicon, consistent with the imaging of the donor wavefunctions. Our results have strong parallels with the observations of deep acceptor states in GaAs. We include in our interpretation a discussion of the varying charge states of the As atoms. The work is of importance to those pursuing the fabrication of devices from individual dopant atoms in silicon, such as quantum computers, where a detailed theoretical understanding of the local properties of the dopants needs to be developed.

## 2. Experiment and simulations

Experiments were performed using an Omicron LT-STM system operated at 77 K with a base pressure below $5 \times 10^{-11}$ mbar. STM tips were prepared by electrochemically etching 0.25 mm diameter polycrystalline W wire before loading into the vacuum system, followed by in-vacuum electron-beam annealing and field emission. Samples were commercial Si(001) wafers (Compart Technology Ltd.), highly doped with arsenic (*n*-type) with the room temperature resistivity of 0.0015-0.004 Ω.cm. Sample preparation involved thorough degassing in UHV at 900 K for 12 hours. This was followed by flash annealing at 1475 K for 10 s. Care was taken to avoid repeated sample flashes that we believe can lead to preferential desorption of n-type dopant atoms from the Si(001) surface region. Hydrogen termination of the Si(001) surface was achieved by exposing the surface to a beam of atomic hydrogen (Tectra atomic hydrogen source) for 5 min, while maintaining the temperature of the sample at 675 K.

We used density functional theory with the Perdew-Burke-Ernzerhof (PBE) [22] functional as implemented in the VASP 4.6.34 code [23]. The core electrons were described by the projector augmented-wave method (PAW) [24], the plane-wave cutoff was 200 eV. All calculations were spin-polarized. We used a (2 x 2 x 1) Monkhorst-Pack [25] k-point mesh. Gaussian smearing of 0.1 eV was used for fractional occupancies. STM images were simulated using the Tersoff-Hamann approach [26]. The computational cell contained four dimer rows with eight dimers per row. The slab consisted of 10 Si layers, terminated by H atoms on both surfaces. We added one electron to the simulation cell of each system to approximately model the degenerately-doped semiconductor.

## 3. Results and discussion

Figure 1(a) shows a filled-state STM image of the Si(001):H surface of an As doped silicon wafer obtained at 77 K. The surface consists of rows of paired silicon atoms called dimers, with each Si atom terminated by a single hydrogen atom. The bright protrusions seen in the filled-state image are dangling bonds (DB) [27-29] that result whenever one of the surface Si atoms is missing its terminating H atom. The features labelled as As1 have not been identified before. These features have the lateral extent of a few nanometres and appear as protrusions superimposed on the background Si(001):H lattice, having a cross-like shape. As figure 1 shows, different As1 features can have different intensities within the same STM image. In empty state images the As1 features all exhibit long-range circular protrusions as we will show in figure 2.

In our STM images of the As-doped Si(001):H surface prepared with minimal thermal treatment as described above, we also commonly observe a second class of features that we label As2. Filled- and empty-state images of this feature are shown in figure 1(b) and 1(c) respectively. In filled state images the As2 feature appears as a circular protrusion with a diameter of a few nanometres. In empty state images the As2 features are seen as an isotropic depression, figure 1(c). The observation of these features forming depressions in empty-state images makes them qualitatively different to all other dopant-related features that we have observed. We discuss the As2 features later in the paper.

The features labeled as As1 in figure 1 all have similar characteristics to one another. Figure 2 represents a more comprehensive characterization of this class of features, showing STM images at several different negative (figure 2(a)) and positive (figure 2(b)) sample biases, probing filled and empty states respectively. All images were obtained with a tunnelling current of 20 pA and are displayed with the same intensity (height) scale. While all of the As1 features appear as protrusions that have a lateral extent of a few nanometres, and are superimposed on the Si(001):H background lattice, their characteristic shapes in filled- and empty-state images are different. In filled state images, these features appear as a number of different shapes, but with certain similarities: the features exhibit maximum intensities along crystallographic directions, producing an appearance that is rectangular, cross-like, or a combination of the two. In addition, there is often structure within the features with certain surface atoms appearing brighter than their neighbours. A good example of this is feature As1.5, see figure 2(a), where the brightest atoms are aligned along the dimer row direction, separated by one surface dimer along the row and by one row perpendicular to the dimer row direction. The intensity of the protrusions is strongly dependent on magnitude of the negative sample bias with the intensity decreasing as the sample bias becomes more negative, and the feature becoming almost indistinguishable from the background surface at -1.5 V.

The appearance of the As1 features in empty-state images (positive sample bias) differs from that of the filled-state images in a two distinct ways. Firstly the features are more isotropic, appearing approximately circular. Secondly, the contrast is fairly uniform across the features with no atoms significantly brighter than their neighbours within the features. However, one similarity with the filled state images is that the intensity of the empty-state features also decreases with increasing bias magnitude.

The bias dependence of the As1 features can be explained following the Mn GaAs literature [16]; the unperturbed As donors are in the electrically neutral ($D^0$) state and are imaged at low negative sample bias when the tip Fermi level is resonant with or below the valence band edge of the sample (figure 2(d)). The As atoms appear as anisotropic protrusions with different characteristic shapes reflecting different projections of the wavefunction of the neutral As atoms on the surface, depending on their precise location in the lattice. At higher filled-state biases the valence-band states dominate tunnelling and eventually render the As atoms almost invisible (figure 2(c)). At positive sample bias shown in figure 2(b), the As atoms are electrically ionized by the tip-induced band bending (TIBB) effect, such that they appear as circular protrusions, reflecting the influence of screened Coulomb potentials of the positively-charged donors ($D^+$) on the conduction band as illustrated in figure 2(e). These circular protrusions gradually disappear at higher positive bias when the tunnelling from the conduction band becomes larger (figure 2(f)).

In order to learn more about the spatial profile of the local density of states (LDOS), we have performed current imaging tunnelling spectroscopy (CITS) on As1.2 and As1.4 features. Figure 3(c) and 3(e) show a two dimensional (2D) cut through these datasets at a constant voltage of 0.95 V over the location of the As1.2 feature, while Figure 3(g) and 3(i) show the equivalent cut over the location of the As1.4 feature. Figures 3(a) and 3(b) show plots of conductance ($dI/dV$) measured as a function of sample bias acquired with the STM tip positioned over the location of an (a) As1.2 feature and (b) As1.4 feature (red traces), as compared to a measurement on a region of the hydrogen-terminated surface (black traces). For the As1.2 and As1.4 features, the appearance of the dopant spectra are qualitatively similar in that there is an enhancement in conductance in the sample voltage range between -1.3 V and -0.6 V. In the case of positive voltages, both dopant spectra show a shift of the onset of tunnelling to lower voltages.

Figures 3(d) and (f), show spatially resolved STS along the line between points 1 and 2 of figure 3(c) (perpendicular to the dimer rows) and points 1 and 2 of figure 3(e) (parallel to the dimer rows), respectively. Two notable effects related to the presence of the donor are seen in the spectra. The first is a downward curving of the spectra at positive voltage with the centre of curvature aligned with the position of the donor. This effect is caused by the screened Coulomb potential created by the positively-charged donor pulling the surface energy levels downwards, resulting in an enhanced tunnelling current and the shift of the onset of tunnelling into the bandgap or to lower positive bias. It is the same effect that causes the circularly isotropic protrusions seen in the STM topographic images figure 2(b). The second effect relates to the spectral aspects at negative voltage where there is no significant downward curvature, implying that there is not the long-range electrostatic effect that would be expected if the dopant were charged. Nevertheless, there is a localized region of increased conductance in the surface bandgap, in the vicinity of the neutral As atom, indicating an energetic state of the dopant inside the bandgap. The nature of the spectra shown in figure 3(c)-(f) for an As1.2 feature are qualitatively reproduced in the spectra of figure 3(g)-(j) for an As1.4 feature. This similarity implies that they have a similar integrated local density of states (LDOS), despite the difference in appearance in the filled-state topographic images shown in figure 2(a). This is consistent with our interpretation that the As1 features reported here derive from As donors that are electrically neutral during filled-state imaging and positively ionized (due to tip induced band bending) for the empty-state images.

To shed light on the nature of the neutral As donors, we have used density functional theory (DFT) modelling to generate simulated STM images of donors in different substitutional lattice sites in the first 4 atomic layers below the surface. We have found that there are two states lying just below the conduction band edge which hybridize with the surface structure to create the cross-like features seen in figures 1 and 2(a). These states are found for As dopants in the third layer of the sample in positions D3 and T3 highlighted in figure 4(a), and are occupied because the samples are degenerately-doped. Simulated STM images of the As dopants in positions D3 and T3 are shown in figures 4(b) and 4(c), respectively.

We now discuss the As2 feature first introduced in figure 1. Three of the As2 features observed are shown again in figure 5, along with two As1 features. The As2 features differ from As1 features in both filled and empty state images. In filled-state images, see figure 5(a), the As2 features appear as circularly isotropic protrusions reminiscent of the As1 appearance in empty-state images. In contrast, when the As2 features are imaged in the empty-state tunnelling, figure 5(b), they appear as circularly isotropic depressions. The contrast behaviour of the As2 feature does not fit the neutral donor model described above because of its circular depression appearance in empty-state images, and moreover it is not consistent with that of positively-charged thermally-ionized donors where we would expect circular protrusions both in filled- and empty-state imaging [20].

It has been previously reported that the electronic features induced by a negatively charged point defect below the Si(001):H surface, such as a thermally-ionized B acceptor, shows the same contrast reversal in filled/empty state images as we see for the As2 feature [19, 20]. (We note that B impurities cannot account for our As2 features based on other STM studies using this wafer.) Therefore we suggest that we might be observing negatively-charged As donors in figure 5. The model proposed in Ref. 19, for B dopants in Si, fits well for explaining the appearance of the As2 feature: when the sample bias is negative, the tip causes downward TIBB as depicted by the solid lines in figure 6(a). The Coulomb potential of a negatively-charged donor locally reduces the downward TIBB, depicted by the dashed lines in figure 6(a), increasing the energy window for tunnelling of the valence band. Hence, with more valence-band states available for tunnelling, the dopant-disturbed surface region appears as a circular protrusion superimposed on the lattice corrugation of the surface. In contrast, when the sample bias is positive, the same Coulomb potential enhances the upward TIBB illustrated in figure 6(b). The energy window for tunnelling of the conduction band is narrowed; therefore, the surface area near the core of the negatively-charged donor appears as circular depression due to the local reduction in tunnelling current compared to the dopant-free surface region.

Additional theoretical and experimental investigations will be necessary to determine why the As donor atoms should become negatively charged in our measurements. We note that there are several reports of DBs at the Si(001):H surface becoming negatively charged owing to the non-equilibrium charging arising from the formation of a depletion region at the surface [29, 30]. Furthermore, recent reports indicate that a hydrogenic donor in semiconductor crystal (Si in GaAs) can act as a negatively-charged dopant inducing a delocalized protrusion in filled-state images and depression when the sample bias is positive [31, 32]. Together these observations suggest that it is possible that some of the As donors we observe here at 77 K are negatively charged; however, further work is required to determine the precise mechanism.

## 4. Conclusions

We have presented an investigation of substitutional As dopants buried beneath a hydrogen terminated Si(001) surface using STM topography and spectroscopy. We find two distinct classes of features that we label As1 and As2. The As1 features present anisotropic, cross-shaped protrusions in low bias filled-state images and circular protrusions in empty-state images. The anisotropic filled-state features are consistent with the direct wavefunction imaging of a neutral donor atom, as has been previously described for acceptors in GaAs. The filled-state appearance is consistent with imaging a positively ionized donor. DFT calculations predict that As dopants in the third layer of the sample induce two states lying just below the conduction band edge, which hybridize with the surface structure creating features with the surface symmetry consistent with our STM images. The As2 features present circular features in both bias polarities; however, in empty-state images these are depressions, rather than protrusions, suggesting a negatively-charged feature. The mechanism for the As donor becoming negatively charged in our measurements is the subject of ongoing investigations.

The work presented here is important for the exploitation of dopants in silicon for future quantum information processing and solotronics applications [33]. Our measurements mark an important step towards the development of a more detailed theoretical understanding of the properties of the subsurface dopants, necessary for the precise wavefunction control that these future applications depend upon.

**Acknowledgements**

We acknowledge G. Aeppli, A.J. Fisher and T. Lim for stimulating discussions. This work was supported by EPSRC EP/H026622/1 and EP/H003991/1.

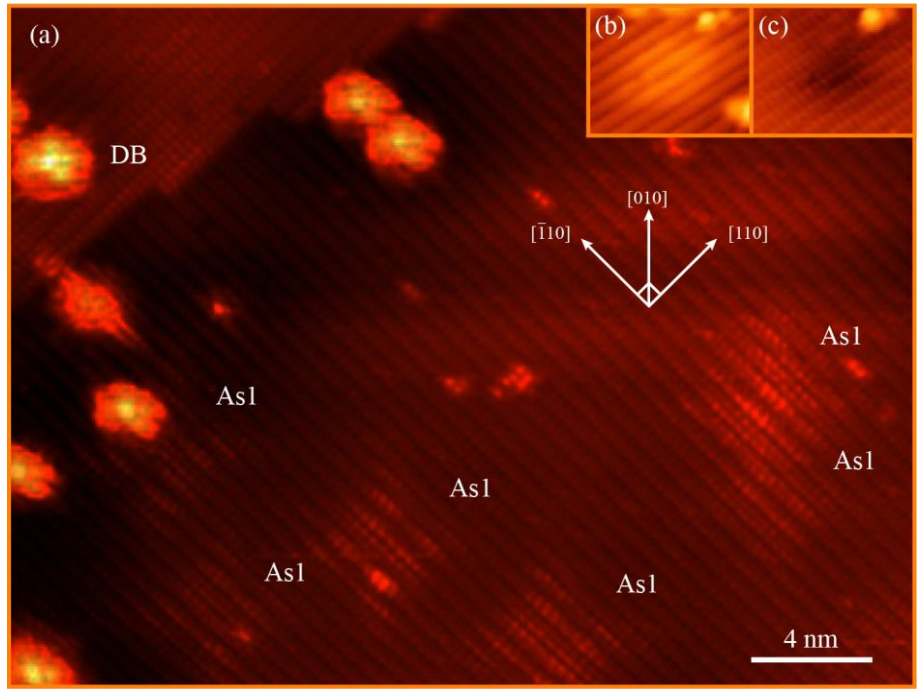

**Figure 1.** A filled-state image of the Si(001):H surface observed by STM at 77 K with a number of As1 features appearing as cross-like protrusions. The image is obtained at sample bias of -1.3 V and tunnelling current of 20 pA. The size of the image is 45 nm (width) x 35 nm (height). (b) and (c) show filled- and empty-state images respectively of an As2 feature obtained at sample bias of -1.3 V and 0.8 V respectively with tunnelling current of 20 pA.

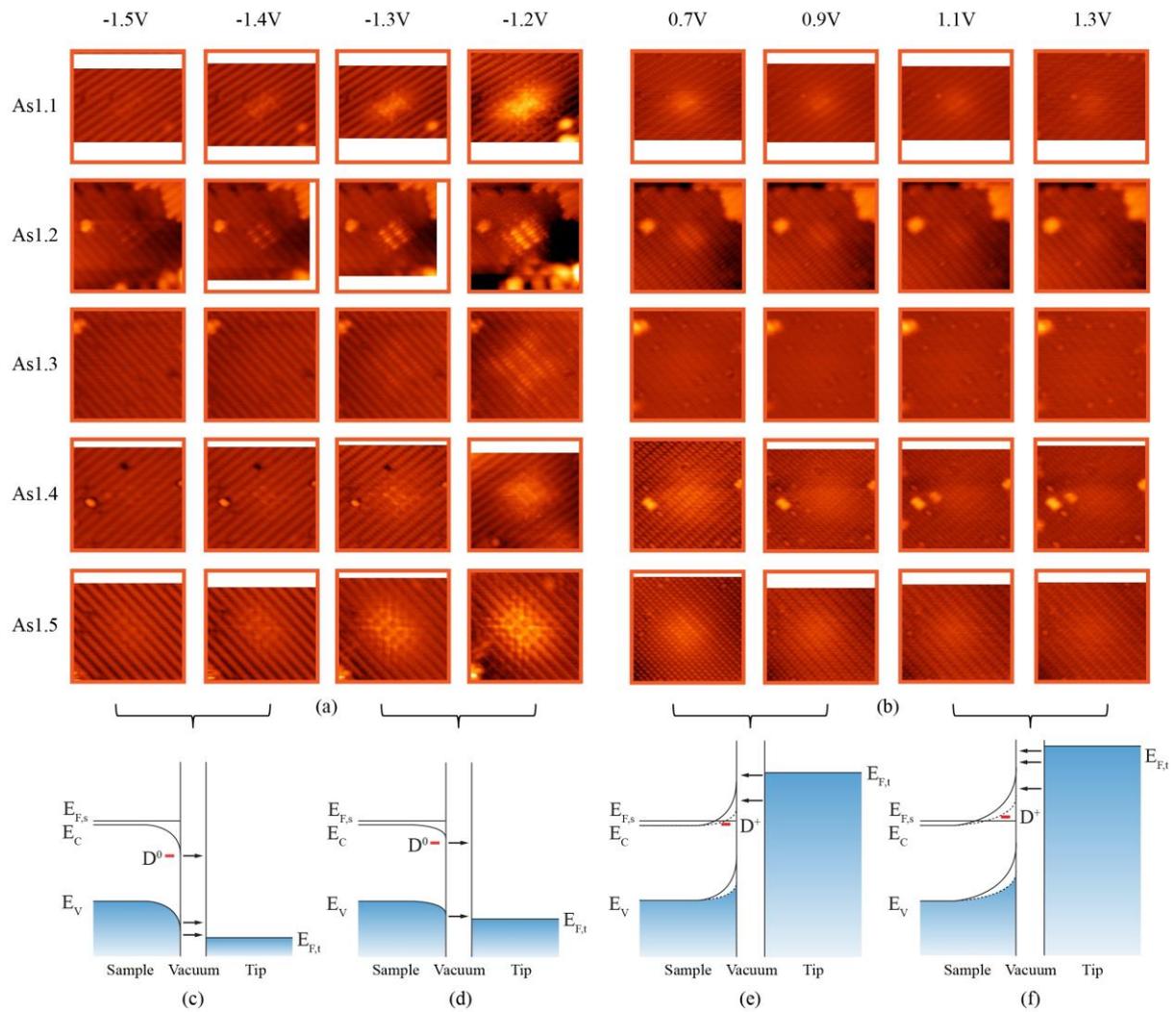

**Figure 2.** (a) Filled- and (b) empty-state images of different As1 features observed by STM at various sample bias with a tunnelling current of 20 pA on the Si(001):H surface at 77 K. Each of the orange boxes encloses a rectangular area of 8 nm x 8 nm. (c-f) Energy-band diagrams describing the tunnelling mechanism at low (d,e) and high voltages (c,f) for both filled- and empty-state imaging regimes.

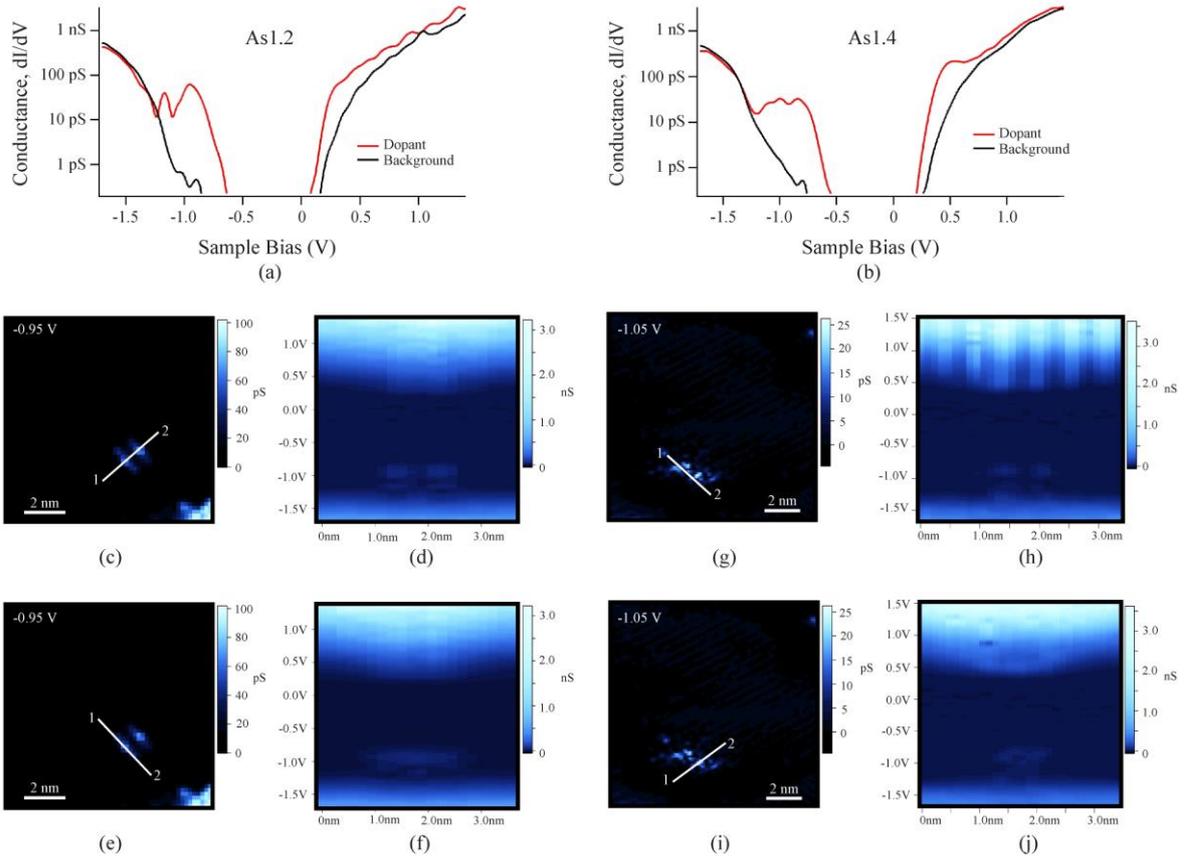

**Figure 3.** (a) and (b) show dI/dV spectra from As1.2 and As1.4 features (red curves) respectively compared to that of the background lattice (black curve). (c-f) CITS data of the As1.2 feature. (g-j) CITS data of the As1.4 feature. In (d), (f), (h) and (j) the colour scale corresponds to conductance plotted as a function of sample bias (vertical axis) and lateral position between the points 1 and 2 (horizontal axis) in the figures (c), (e), (g) and (i) respectively.

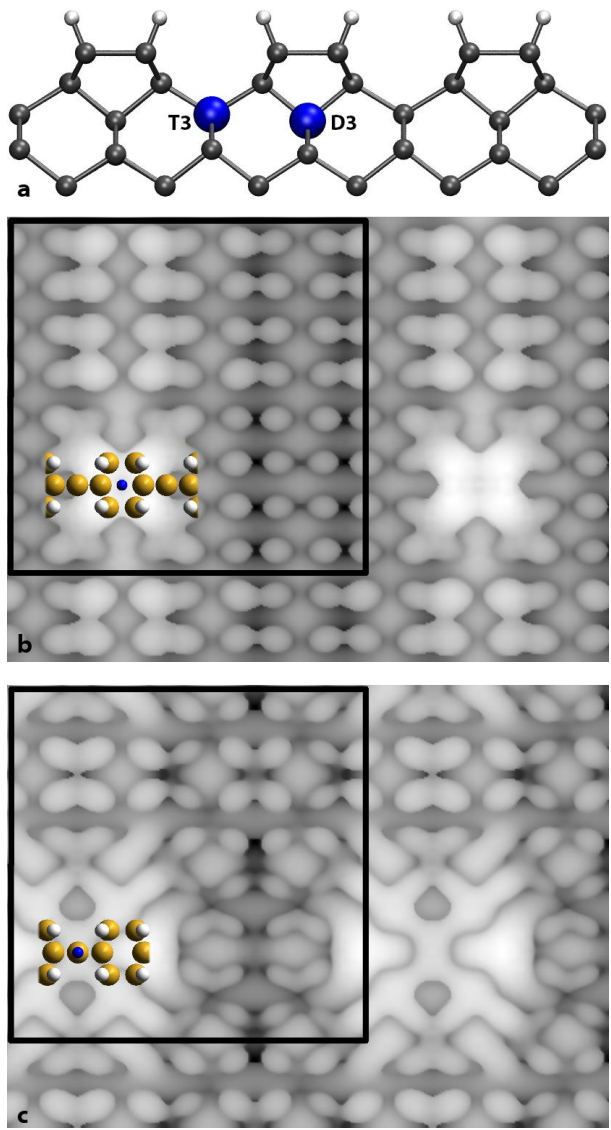

**Figure 4.** (a) Side view of the top layers of the computational slab and the two positions of the dopants that best reproduce experimental data. The Si atoms are grey, H atoms white, dopant atoms blue and large. b) and c) Simulated STM images of the system with a dopant in the D3 (b) and T3 (c) site, respectively. Si atoms are large and yellow, H atoms white and small, dopant atom positions are blue and small.

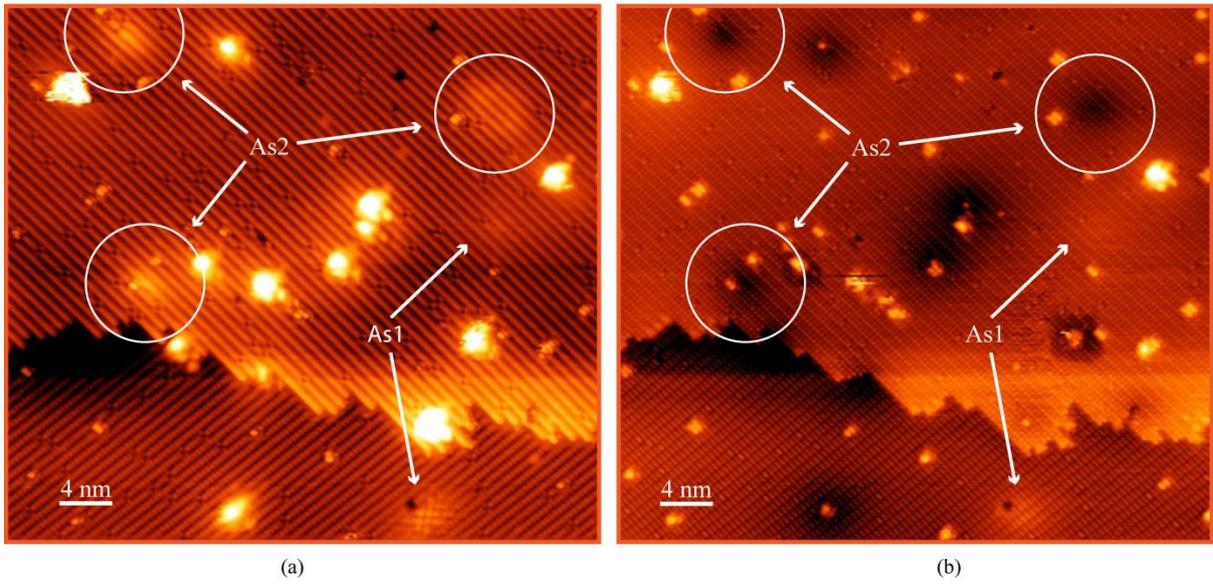

**Figure 5.** (a) Filled-state and (b) empty-state STM image of As1 and As2 features on the Si(001):H surface obtained at sample voltages of -1.3 V and 0.8 V respectively, and a tunnelling current of 20 pA. The size of each image is 40 nm x 35 nm.

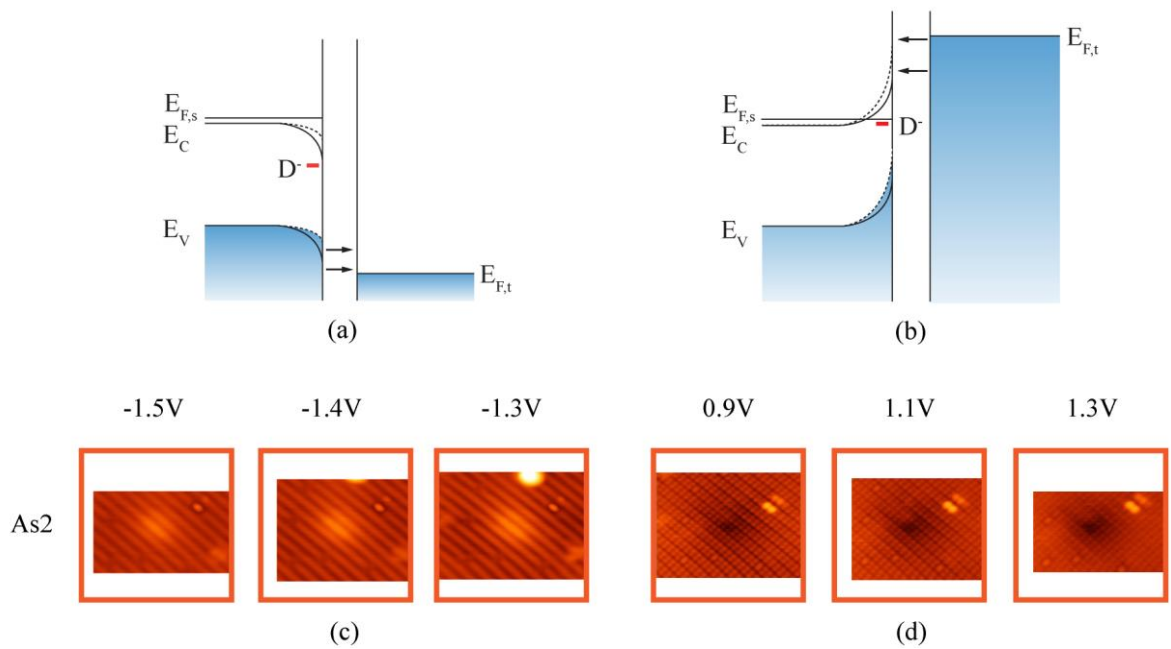

**Figure 6.** (a) and (b) The energy-band diagrams describing tip-induced band bending perturbed by a negatively charged subsurface As donor under filled-state and empty-state tunnelling conditions respectively – see main text for details. (c, d) Voltage-dependent images of an As2 feature showing that the feature become less strongly resolved with increasing bias voltage magnitude for both negative and positive bias. The size of each image is 8 x 8 nm$^2$.